\documentclass{osa-article}
\journal{oe}

\newcommand{\micron}[0]{$~\upmu$m }
\begin{document}

\title{Implementing non-scalar diffraction in Fourier optics via the Braunbek method}

\author{Anthony Harness}

\address{Mechanical \& Aerospace Engineering Department, Princeton University, Princeton, NJ, 08544, USA}

\email{aharness@princeton.edu} 


\begin{abstract}
Fourier optics is a powerful and efficient tool for solving many diffraction problems, but relies on the assumption of scalar diffraction theory and ignores the three-dimensional structure and material properties of the diffracting element. Recent experiments of sub-scale starshade external occulters revealed that the inclusion of these physical properties is necessary to explain the observed diffraction at 10$^{-10}$ of the incident light intensity. Here, we present a methodology for implementing non-scalar diffraction while maintaining the efficiency and ease of standard Fourier optics techniques. Our methodology is based on that of Braunbek, in which the Kirchhoff boundary values are replaced with the exact field in a narrow seam surrounding the edge of the diffracting element. In this paper, we derive the diffraction equations used to implement non-scalar diffraction and outline the computational implementation used to solve those equations. We also provide  experimental results that demonstrate our model can replicate the observational signatures of non-scalar diffraction in sub-scale starshades, in effect validating our model to better than 10$^{-10}$ in relative intensity. We believe this method to be an efficient tool for including additional physics to the models of coronagraphs and other optical systems in which a full electromagnetic solution is intractable.
\end{abstract}

\section{Introduction}
\label{sec:intro}
For many optical systems, appropriate assumptions can be made to enable the use of Fourier optics as a powerful and efficient tool for understanding diffraction and the propagation of light. The most consequential assumption necessary is that all components of the electromagnetic field obey an identical scalar wave equation. This assumption, and the resultant scalar diffraction theory, is valid if propagating in a dielectric medium that is linear, isotropic, homogeneous, and nondispersive\cite{Goodman}. It is also necessary for diffracting elements to be much larger than the wavelength of light to ensure no coupling between field components at the boundaries\cite{Goodman}. For most optical systems, particularly visible wavelength astronomical telescopes, these assumptions are sufficiently valid. The optical models for exoplanet detection instruments such as high contrast coronagraphs\cite{Krist_2007} and starshade external occulters\cite{Cash_2011,Cady_2012,Harness_2018} rely on Fourier optics and the scalar diffraction assumption to predict the performance to $<10^{-10}$ of the incident light. Up to this point, standard physical optics propagation techniques based on Fourier optics have been sufficient for many applications, including starshades. However, recent laboratory experiments with sub-scale starshades have revealed the presence of non-scalar diffraction, a phenomenon which we deem the \emph{thick screen effect}, as light propagates past the thick optical edges that form the narrow gaps between starshade petals\cite{Harness_2020}.

In order to properly validate starshade optical models with sub-scale experiments, the models need to account for the material properties and geometry of a thick diffraction screen, along with polarization effects, which are available through Maxwell's equations. However, resorting to solutions of Maxwell's equations nullifies the efficiency with which Fourier optics operates. It is intractable to directly solve Maxwell's equations over the entirety of the starshade because its overall dimensions are much larger than the wavelength of light (the diameter of a sub-scale starshade is $\sim50,000\lambda$).

Our approach, based on an insight provided by Braunbek\cite{Braunbek}, is to use the solution to Maxwell's equations in a narrow seam around the edge of the diffraction screen and apply those results to the scalar model through standard Fourier optics techniques.  Ref.~\citenum{Meneghini} uses the Braunbek method and Sommerfeld's solution to derive a Maggi-Rubinowicz vector potential for a perfectly conducting screen, but the vector potential cannot be generalized for more complex edge geometries and materials. Our methodology provides that generalization. This technique is necessary to replicate experimental data of sub-scale starshades, but we also believe it to be useful for coronagraphs and other optical systems that possess a large dynamic range in size.

Section~\ref{sec:thick_screen_effect} of this paper introduces the thick screen effect observed in sub-scale starshade experiments that prompted the need for a new modeling technique. Section~\ref{sec:derivation} gives the formal derivation of the model. Section~\ref{sec:computational} summarizes the computational implementation that integrates non-scalar diffraction into the Fourier optics model. Section~\ref{sec:edge_diffraction} outlines the finite-difference time-domain (FDTD) code used to solve Maxwell's equations for realistic optical edges. Section~\ref{sec:experimental} provides experimental data that show our method agrees with experimental results at relative intensity levels of $<10^{-10}$. Section~\ref{sec:conclusions} restates the conclusions drawn from this work.

\section{Thick screen effect in sub-scale starshade experiments}
\label{sec:thick_screen_effect}
Optical experiments of 1/1000$^\mathrm{th}$-scale starshades are being conducted to verify optical performance and validate optical models at flight-like Fresnel numbers\cite{Harness_2020}. Recent results have shown contrast (relative intensity) levels that are orders of magnitude above that predicted by scalar diffraction theory and with a morphology dependent on the input polarization state. This was deemed the thick screen effect\cite{Azpiroz_2003} and was first reported in Ref.~\citenum{Harness_2020}. We briefly summarize the thick screen effect below.

The petalized shape of a starshade is a binary approximation to a radial apodization function that has been numerically optimized to suppress starlight by $>10$ orders of magnitude. The optimization routine\cite{Vanderbei_2007} that solves for the apodization function assumes an infinitely thin optical edge and that all features of the starshade are much larger than the wavelength of light and thus the scalar diffraction assumption holds. This assumption is likely to be valid for large starshades in a space mission where the minimum feature size is on the order of millimeters, however, the starshades tested in the laboratory are 1000$\times$ smaller and have minimum feature sizes on the order of microns. Once starshade experiments reached 10$^{-10}$ contrast levels at a low Fresnel number, the presence of bright polarization-dependent lobes revealed a break down in scalar diffraction theory\cite{Harness_2020}.

\begin{figure}[htb]
\centering\includegraphics[width=0.9\textwidth]{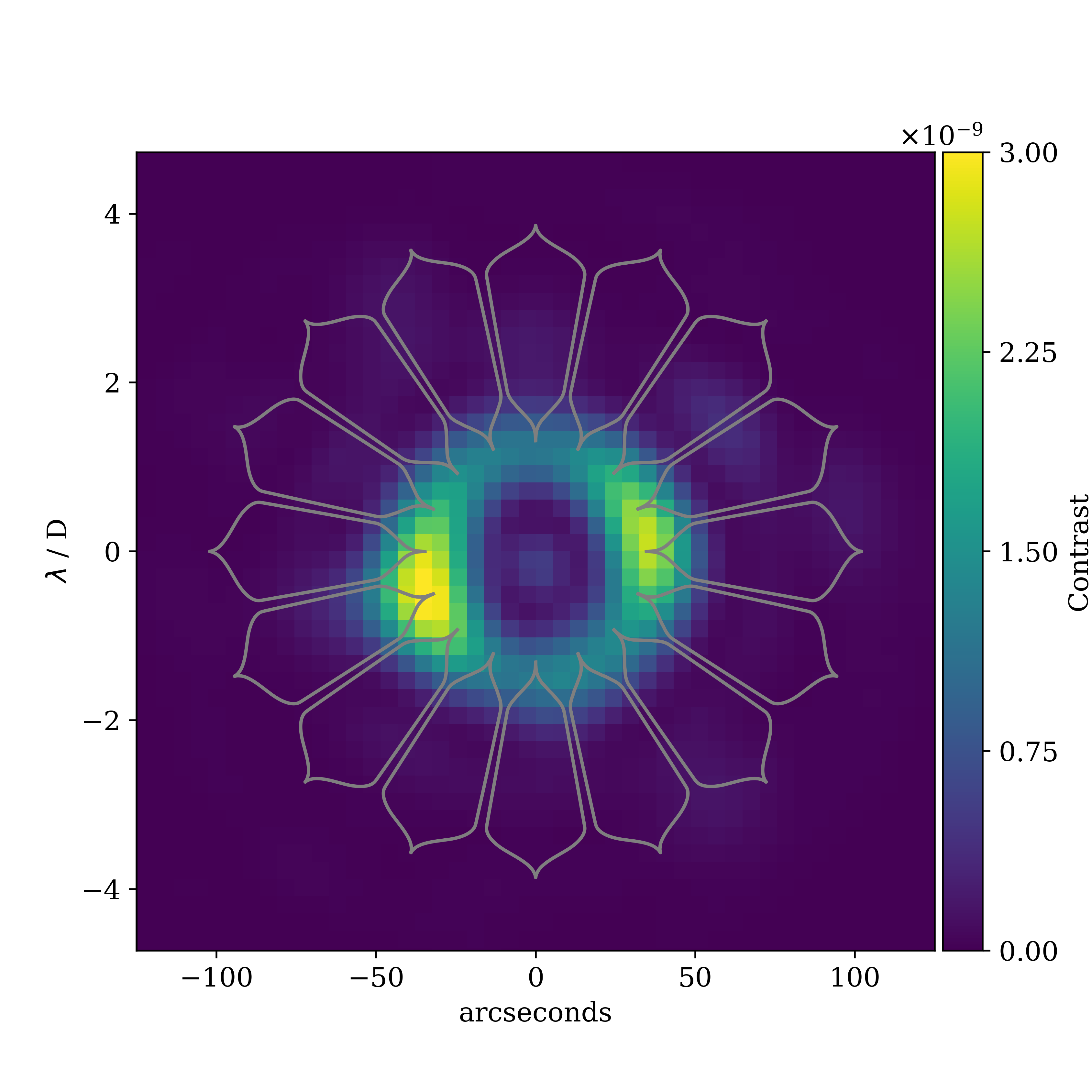}
\caption{ \label{fig:lab_image}
Experimental contrast image of mask DW9 at $\lambda$ = 641 nm. The starshade pattern is overlaid.
}
\end{figure}

The starshades tested in these experiments are 25 mm in diameter and etched out of silicon-on-insulator (SOI) wafers. The gap width between starshade petals is as small as 7.5\micron and the device layer of the SOI wafer that serves as the optical edge is as thick as 7~$\upmu$m, meaning the edge can no longer be considered a thin screen as required in the Fresnel-Kirchhoff diffraction equation. As light propagates through the narrow gaps between petals, and past the thick screen, it interacts in a polarization-dependent manner with the material properties of the edge that induce a slight change in the electric field (in both amplitude and phase). The change in the field translates to a change in the effective apodization profile of the petal and negates some of the optimized light suppression. The resultant contrast appears as bright lobes that are aligned with the input polarization vector, as can be seen in Fig.~\ref{fig:lab_image} and as will be discussed further in Sec.~\ref{sec:experimental}.

As Braunbek\cite{Braunbek} and Keller\cite{Keller} have pointed out, and as the FDTD simulations in Sec.~\ref{sec:edge_diffraction} show, diffraction is a local phenomenon affecting only the immediate surrounding of the edge (akin to a boundary layer). Interaction with the edge is present in all optical systems, but for those that have apertures much larger than the wavelength, the contribution from a few $\lambda$-wide boundary layer is negligible. For the $\sim10\lambda$ wide gaps in the sub-scale starshades, the boundary layer occupies an appreciable fraction of the aperture and thus it has a significant effect, especially when the shape of that aperture is precisely designed to suppress the light intensity by 10 orders of magnitude. This explanation is supported by the degradation in contrast only appearing at the narrow inner gaps between petals, where the fraction of the aperture occupied by the $\lambda$-wide boundary layer is greatest. The degradation in contrast is not seen in the gaps at the outer regions of the mask because at larger radii the Fresnel number is larger and thus the sensitivity to changes in the apodization profile is lower. This physical description of the thick screen effect helps inform the derivation of the model in the following section.

\section{Model derivation}
\label{sec:derivation}
In the derivation of the Fresnel-Kirchhoff diffraction equation, it is assumed that the field in the aperture of a diffraction screen is not affected by the presence of the screen. Thus in the plane of the screen, one can assign Kirchhoff's boundary values, where the field is zero on the screen and equal to the incident field ($U_0$) in the aperture\cite{Born_Wolf}. The main insight introduced by Braunbek was to replace the boundary values of Kirchhoff with the exact value of the field in a narrow seam (a few $\lambda$'s wide) around the screen's edge\cite{Braunbek}. Braunbek obtained the exact value in the seam from Sommerfeld's solution to the perfectly conducting half-plane. The approach of this work is to adapt Braunbek's insight to replace the boundary values in the seam around the edge, with the exact field that results from the presence of the thick screen. The exact field in the presence of an arbitrary screen can be calculated via numerical solution of Maxwell's equations, as will be demonstrated in Sec.~\ref{sec:edge_diffraction}.

\begin{figure}[htb]
\centering\includegraphics[width=7cm]{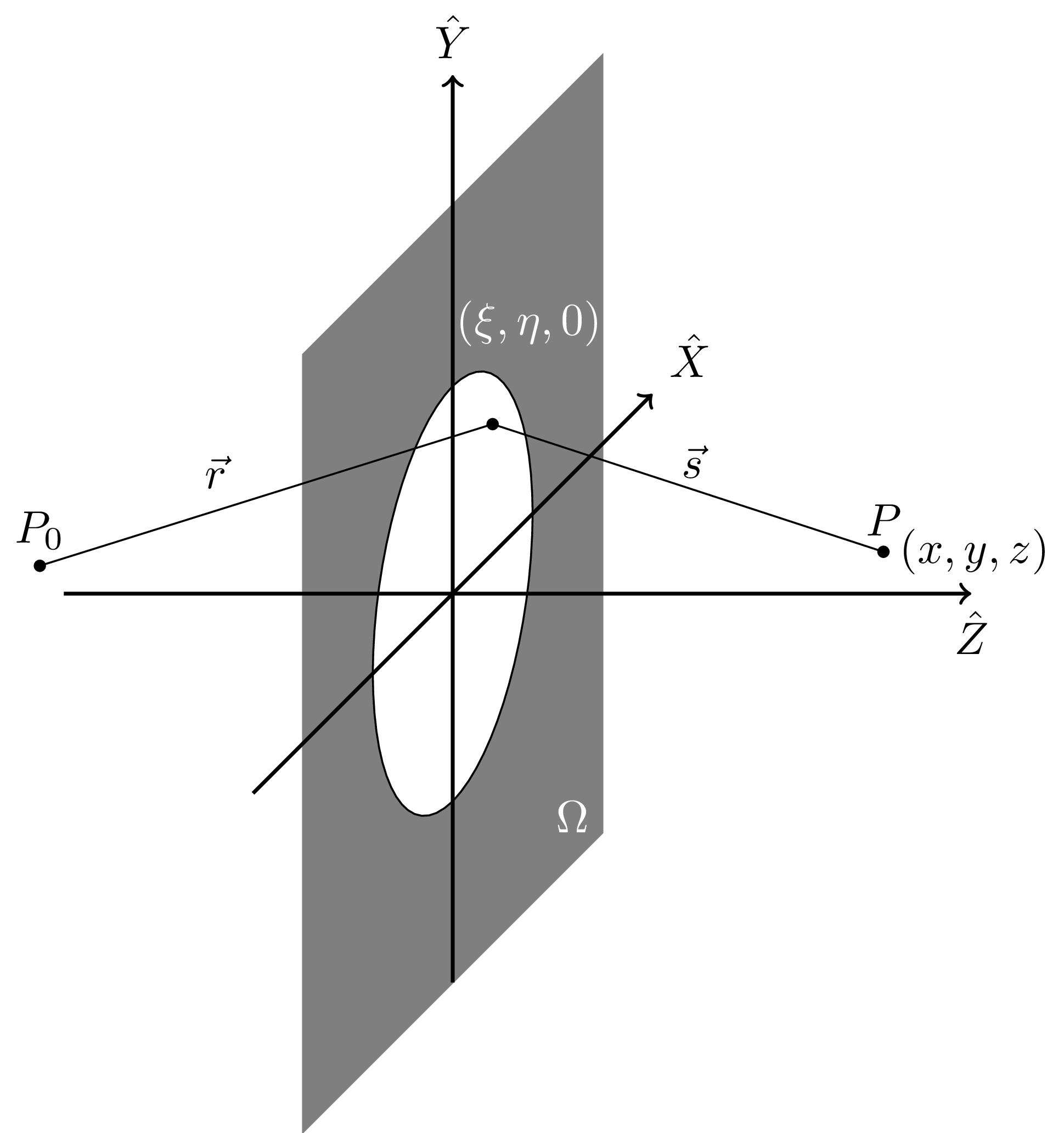}
\caption{ \label{fig:lab_frame}
Definition of the lab coordinate system used for the derivation of the model. The source is located at $P_0$, the diffraction screen is located flat in the $z=0$ plane, the integral is taken over the entire $z=0$ surface $\Omega$, and the observation point is at $P$.
}
\end{figure}

We assume that the electromagnetic field can be described by a scalar potential $U$; later when polarization is taken into account, we will assume that this holds for each component of the field. Starting with the integral theorem of Helmholtz and Kirchhoff\cite{Born_Wolf}, the disturbance at a point $P$ is given by
\begin{equation}
	U\left(P\right) = \frac{1}{4\pi} \iint_\Omega\left(U\frac{\partial G}{\partial n} - \frac{\partial U}{\partial n}G\right) \, dA \,,
	\label{eq:helmholtz_kirchoff}
\end{equation}
where $\partial/\partial n$ is differentiation normal to the screen and both $U$ and the Green's function $G$ satisfy the Helmholtz equation. The integral is taken over the entire surface $\Omega$ (see Fig.~\ref{fig:lab_frame}). We assume the screen is flat in the $z=0$ plane such that $\partial/\partial n = \partial/\partial z$.

\begin{figure}[htb]
\centering\includegraphics[width=7cm]{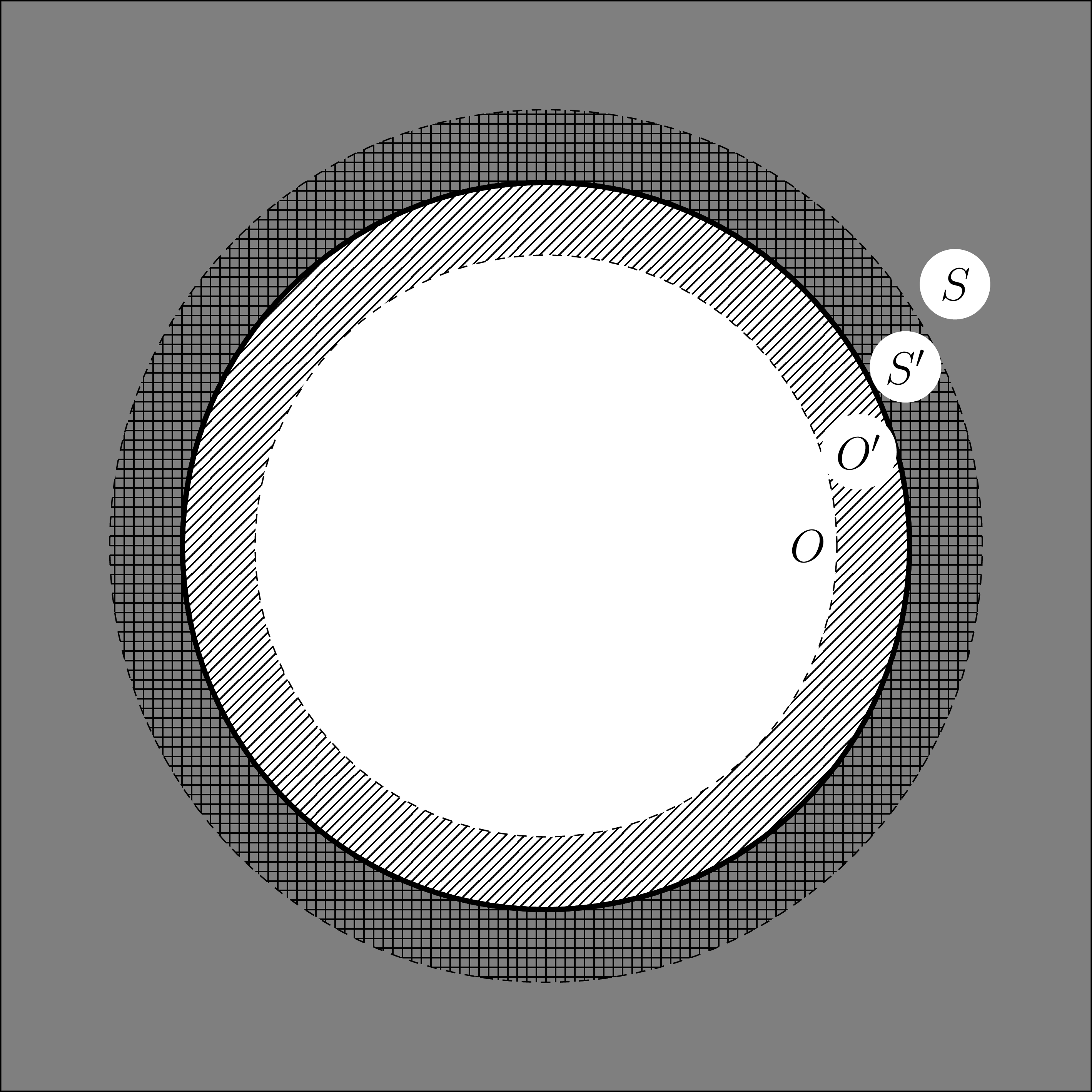}
\caption{ \label{fig:screen_areas}
The surface $\Omega$ is divided into four regions: the screen $S$, the screen side of the seam $S^\prime$, the aperture side of the seam $O^\prime$, and the aperture $O$. The screen edge is the solid black line between the $S^\prime$ and $O^\prime$ surfaces. The size of the seam is exaggerated for clarity.
}
\end{figure}

Building on Braunbek's methodology, we split the entire surface into four distinct regions: $\Omega = S \cup S^\prime \cup O^\prime \cup O$, where $S$ is the screen (excluding the seam), $O$ is the aperture (excluding the seam), $S^\prime$ is the screen side of the seam, and $O^\prime$ is the aperture side of the seam (see Fig.~\ref{fig:screen_areas}). We assume the field in the seam is a superposition of the incident field and an additional field $\delta$, which is induced by the presence of the screen. In other words, the boundary values on a surface $A$ are given by
\begin{equation}
    U = \begin{cases}
        0, & A \subseteq S\\
        \delta,  & A \subseteq S^\prime\\
        U_0 + \delta, & A \subseteq O^\prime\\
        U_0, & A \subseteq O\,.\\
    \end{cases}
    \label{eq:boundary_values}
\end{equation}

The additive field $\left(\delta\right)$ can be considered a better approximation to the true boundary values near the edge of the screen. Its existence can be physically understood as due to a boundary diffraction wave emanating from the edge of the screen\cite{Rubinowicz}, but $\delta$ is not formally equivalent to the boundary diffraction wave. Although Kirchhoff's solution can be written as the sum of the incident wave and the boundary diffraction wave\cite{Rubinowicz}, that analysis still assumes Kirchhoff's boundary values in which $\delta=0$.

Substituting Eq.~(\ref{eq:boundary_values}) into Eq.~(\ref{eq:helmholtz_kirchoff}), the Helmholtz-Kirchhoff integral can be split into two integrals,
\begin{equation}
    U\left(P\right) = U_K\left(P\right) + U_B\left(P\right) \,,
\end{equation}
with one term representing the standard Kirchhoff diffraction integral:
\begin{equation}
	U_K\left(P\right) = \frac{1}{4\pi} \iint_{O^\prime+O}\left(U_0\frac{\partial G}{\partial z} - \frac{\partial U_0}{\partial z}G\right) \, dA \,,
	\label{eq:kirchhoff_integral}
\end{equation}
and an additional term, deemed the Braunbek integral, due to the additional field in the seam:
\begin{equation}
	U_B\left(P\right) = \frac{1}{4\pi} \iint_{S^\prime + O^\prime}\left(\delta\frac{\partial G}{\partial z} - \frac{\partial \delta}{\partial z}G\right) \, dA \,.
	\label{eq:braunbek_integral}
\end{equation}

Note that the two integrals are of identical form, but use different boundary values and are taken over different domains (see Eq.~(\ref{eq:boundary_values})). The Kirchhoff solution is calculated over the entire aperture (including the seam), but not over the screen as the incident field on the screen is zero. The Braunbek solution is calculated over the entire seam. The surface $O^\prime$ is integrated over by both solutions, as $O^\prime$ sees both the incident field and the additive Braunbek field.

From here, we can follow standard practice to simplify the integrals to a more usable form. We make the paraxial approximation and assume the distances from the source and the observation point to the screen are much larger than the wavelength. We assume a spherical incident wave, which we write as
\begin{equation}
    U_0 = \frac{e^{ikr}}{r}, \frac{\partial U_0}{\partial z} \approx ik\frac{e^{ikr}}{r} = ikU_0\,,
\end{equation}
and write the Green's function on the observation side as
\begin{equation}
    G = \frac{e^{iks}}{s}, \frac{\partial G}{\partial z} \approx -ik\frac{e^{iks}}{s} \,.
    \label{eq:greens_function}
\end{equation}

If the Fresnel approximation is made on distances $r$ and $s$, keeping the first order term in amplitude and second order term in phase, we can write the Kirchhoff integral as the Fresnel diffraction equation:
\begin{equation}
    U_K\left(x,y,z\right) = \frac{e^{ikz}}{i\lambda z} \iint_{O^\prime + O} U_0
    e^{\frac{ik}{2z}\left[\left(\xi-x\right)^2 + \left(\eta-y\right)^2\right]}  \, d\xi \, d\eta \,.
    \label{eq:K_fresnel}
\end{equation}

\begin{figure}[htb]
\centering\includegraphics[width=10cm]{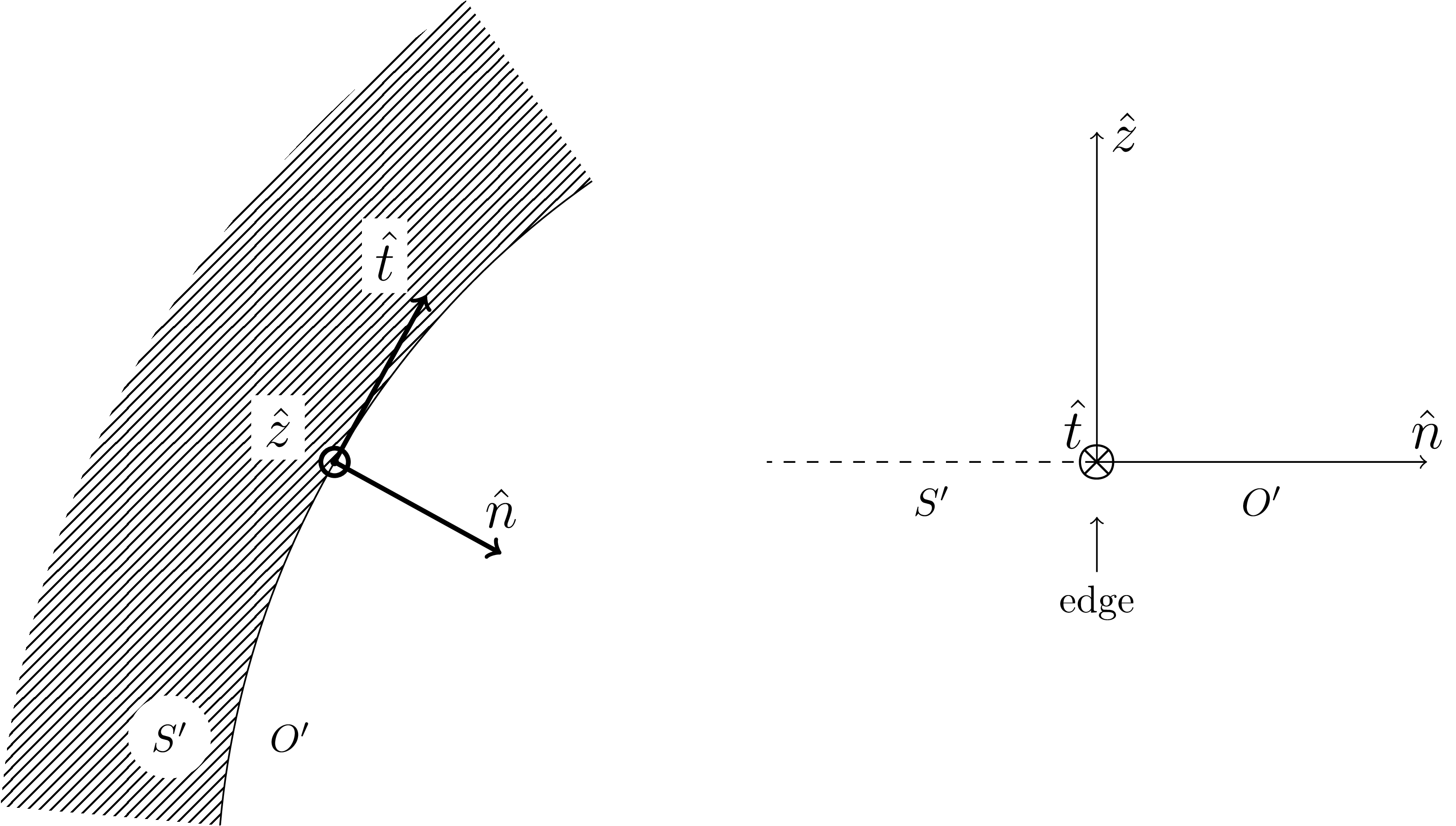}
\caption{ \label{fig:edge_frame}
Definition of the edge coordinate system centered at each edge point. $\hat{n}$ is normal to the edge, $\hat{t}$ is tangent to the edge and points clockwise, and $\hat{z}$ points towards the observer. Left: front facing view from the observation point. Right: view in the plane of the screen looking clockwise along the edge.
}
\end{figure}

We can follow a similar derivation for the Braunbek integral, but first need to transition from a purely scalar description of the electromagnetic field to one that takes polarization into account. We define an edge coordinate frame in Fig.~\ref{fig:edge_frame} and assume that the screen is in the plane orthogonal to $\hat{z}$ so that $\hat{z}_\mathrm{edge}=\hat{z}$; $\hat{n}$ is normal to the edge in the plane of the screen (this is different from $n$ in Eq.~(\ref{eq:helmholtz_kirchoff})); and $\hat{t}$ is tangent to the edge in the clockwise direction. For each point on the edge, the edge is treated as an infinite half-plane (the radii of curvature of the edges are $>>\lambda$) making the field independent of $\hat{t}$. Our problem is now two-dimensional and the field can be described in terms of a single dependent variable, which behaves in a scalar fashion\cite{Born_Wolf}.  We break the field into two independent polarization states: $s$-polarization, where the electric field is parallel to the edge and the complete field is described by $E_t$; and $p$-polarization, where the magnetic field is parallel to the edge and the field is described by $H_t$. We can now apply the same derivation of the Kirchhoff integral to $E_t$ and $H_t$.

\begin{table}[ht]
\caption{Electric and magnetic field values taken by the additive field for a given polarization.}
\label{tab:delta_values}
\begin{center}
\begin{tabular}{ c | c | c }
\hline
\rule[-1ex]{0pt}{3.5ex} Polarization & $\delta$ & $\partial\delta/\partial z$\\
\hline\hline
\rule[-1ex]{0pt}{3.5ex} s & $E_t$ & $\partial E_t/\partial z = -ikH_n$\\
\rule[-1ex]{0pt}{3.5ex} p & $H_t$ & $\partial H_t/\partial z = ikE_n$\\
\hline
\end{tabular}
\end{center}
\end{table}

In Table~\ref{tab:delta_values}, we relate $\partial \delta/\partial z$ to field components using Maxwell's equations in free space. Inserting these values and those of Eq.~(\ref{eq:greens_function}) into Eq.~(\ref{eq:braunbek_integral}), and making the Fresnel approximation, we write the Braunbek integral for $s$- and $p$-polarization as
\begin{equation}
    U_{B,\begin{smallmatrix}s\\p\end{smallmatrix}}\left(x,y,z\right) = \frac{e^{ikz}}{i\lambda z}  \iint_{S^\prime + O^\prime} \left(E_{\begin{smallmatrix}t\\n\end{smallmatrix}} \mp H_{\begin{smallmatrix}n\\t\end{smallmatrix}}\right)
    e^{\frac{ik}{2z}\left[\left(\xi-x\right)^2 + \left(\eta-y\right)^2\right]}  \, d\xi \, d\eta \,.
    \label{eq:B_fresnel}
\end{equation}

Again, the diffraction equations for the Kirchhoff and Braunbek solutions are the same, but with different initial field values on the surfaces. We will calculate both integrals in the same manner, so we write both integrals in the general form,
\begin{equation}
    U_\gamma\left(x,y,z\right) = \frac{e^{ikz}}{i\lambda z} \iint_{A_\gamma} P_\gamma
    e^{\frac{ik}{2z}\left[\left(\xi-x\right)^2 + \left(\eta-y\right)^2\right]}  \, d\xi \, d\eta \,.
    \label{eq:general_diff}
 \end{equation}
where $\gamma = \left\{K,s,p\right\}$ indicates the solution (Kirchhoff; $s,p$ polarizations of Braunbek) and $P_\gamma$ is the corresponding initial field over the surface $A_\gamma$, which also takes into account the aperture function. $P_\gamma$ is described in Sec.~\ref{sec:greypixel}. Eq.~(\ref{eq:general_diff}) can be simplified as the Fourier transform (represented as $\mathcal{F\{\}}$) of the initial field times a Fresnel kernel:
\begin{equation}
    U_\gamma\left(x,y,z\right) = \frac{e^{ikz}}{i\lambda z} e^{\frac{ik}{2z}\left(x^2+y^2\right)} \mathcal{F}\left\{P_\gamma e^{\frac{ik}{2z}\left(\xi^2 + \eta^2\right)}\right\}_{\left(x,y\right)} \,.
    \label{eq:general_diff_fourier}
 \end{equation}

This shows that both the Kirchhoff and Braunbek solutions can be efficiently calculated with standard Fourier techniques.

\section{Computational implementation}
\label{sec:computational}
In Eq.~(\ref{eq:general_diff_fourier}), we formulated the diffraction equation as the Fourier transform of a Fresnel kernel times a map of the initial field values ($P_\gamma$) on the surface $(A_\gamma)$, whose value depends on the solution being calculated $\left(\gamma\right)$ and whether or not the surface is inside the seam $\left(A_\gamma\subseteq S^\prime\cup O^\prime\right)$. Braunbek posited the seam to be a few $\lambda$'s wide\cite{Braunbek}, after which the field approaches the value of the incident field. We select the width to be that in which the contrast solution is numerically converged, typically $\sim$ 10\micron total seam width for the experiment described in Sec.~\ref{sec:experimental}. This section will describe the construction of $P_\gamma$, the implementation of polarization effects, and the numerical computation of the diffraction equation.

\subsection{Greypixel map of initial field}
\label{sec:greypixel}
To enable the use of efficient Fourier transform algorithms, we represent $P_\gamma$ and the coordinates $\xi,\eta$ as square two-dimensional matrices spanning the integral surface. At the edge of the screen, there is a sharp discontinuity in the binary aperture function where the discretization of the matrix reduces the accuracy of the calculation. Standard practice for minimizing the effect of discretization is to soften the discontinuity by giving cells that lie on the edge a value between 0 and 1, depending on what fraction of the cell lies inside the aperture. We call a matrix with this sub-pixel resolution a ``greypixel map'' and the technique is employed below.

For cells outside the Braunbek seam, only the Kirchhoff solution is used; those in the aperture are given value 1, those on the screen are given value 0. A winding number algorithm\cite{Press_1992} can determine if the cell is inside or outside the aperture.

\begin{figure}[htb]
\centering\includegraphics[width=\textwidth]{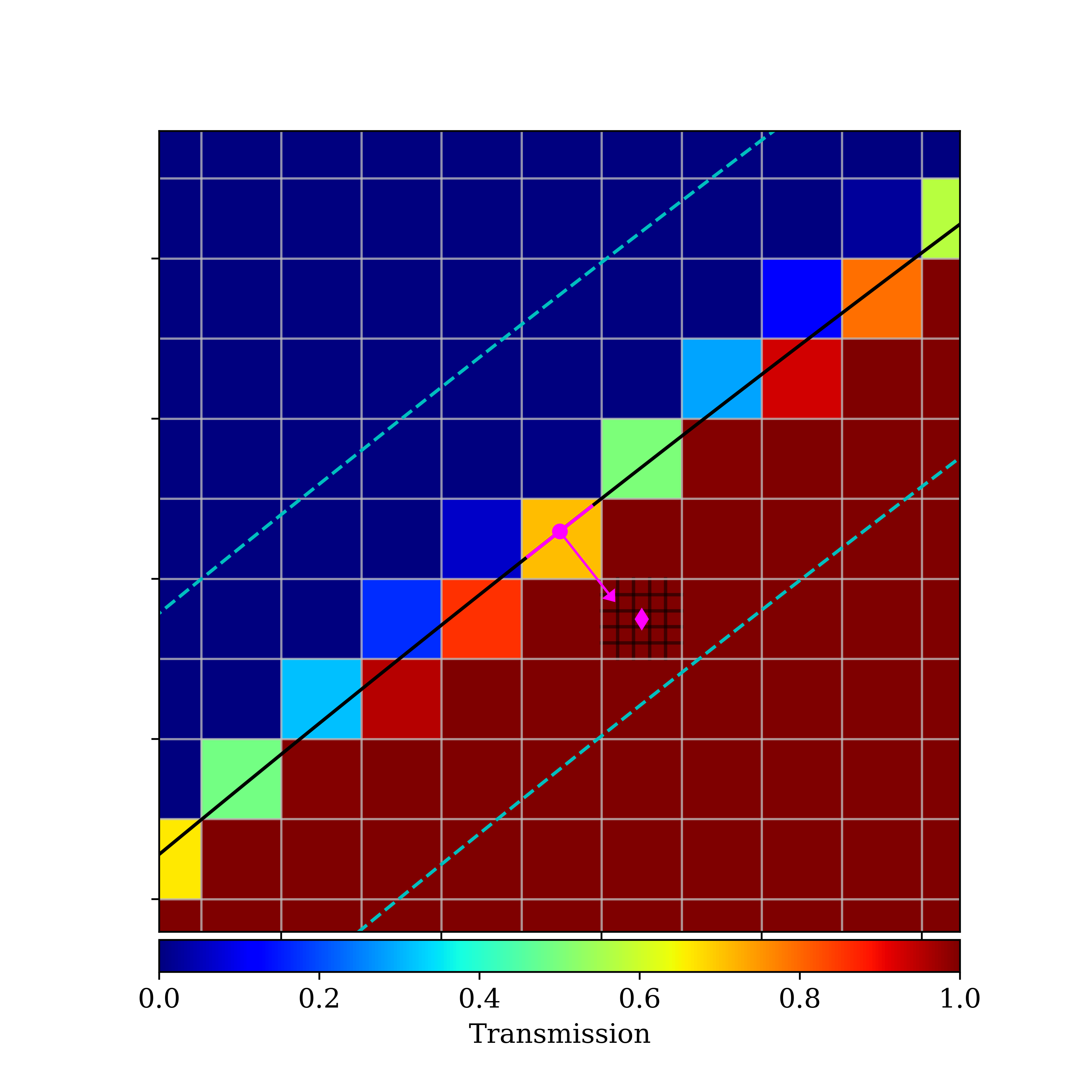}
\caption{ \label{fig:greypixel}
 Example greypixel map showing scalar aperture transmission ($f_K/U_0$). The black line is the aperture edge and the cyan dashed lines mark the 10\micron wide seam. For each cell under consideration (magenta diamond) the nearest edge point (magenta circle) is found and the distance along the normal (magenta arrow) is calculated. The cell is subdivided into subpixels to integrate the field function as in Eq.~(\ref{eq:greypixel_value}).
}
\end{figure}

For each cell in the seam, we find the nearest edge point (see Fig.~\ref{fig:greypixel}) and calculate the implicit line equation for the edge (assuming it has no curvature), along with the angle normal to the edge. We divide the cell into a $N\times N$ sub-pixel grid (with sub-pixel size $\Delta\xi\times\Delta\eta$) and for each sub-pixel, we calculate the distance $d_{i,j}$ projected to the edge segment via the implicit line equation. From the edge distance, we calculate the field value for each sub-pixel and integrate (via the trapezoidal rule) over the cell to give the total field in the cell. Thus the initial field at each cell is given by
\begin{equation}
    P_\gamma\left(\xi,\eta\right) =  \Delta\xi\Delta\eta\sum\limits_{i=0}^{N} \sum\limits_{j=0}^{N} f_\gamma\left(d_{i,j}\right) \,.
    \label{eq:greypixel_value}
\end{equation}

Here, $f_\gamma$ is the initial field due to the edge effect as a function of distance, for solution $\gamma$. For $\gamma=K$, i.e., the scalar Kirchhoff solution, the edge function is represented by the incident field times the Heaviside step function:
\begin{equation}
    f_K(d) = U_0\begin{cases}
        0, & d < 0\\
        1, & d \ge 0 \,.
    \end{cases}
    \label{eq:heaviside}
\end{equation}

The distance $d$ is signed, with negative values on the side of the screen. Inspection of Eq.~(\ref{eq:heaviside}) shows it yields the Kirchhoff boundary values of Eq.~(\ref{eq:boundary_values}): when $d$ is negative (the screen side of the edge), the field is zero; when $d$ is positive (the aperture side of the edge), the field is equal to the incident field. For those cells in which the edge passes through, the aperture function is between 0 and 1, providing the greypixel effect to enhance resolution. Figure~\ref{fig:greypixel} shows an example greypixel map for the Kirchhoff solution.

For the Braunbek solutions, $f_\gamma$ is complex-valued as it accounts for both amplitude and phase. If we assume the screen is thin and perfectly conducting, $f_\gamma$ is given analytically by Sommerfeld's solution to diffraction by a half-plane\cite{Sommerfeld, Born_Wolf}. For more complex edge geometries and materials, $f_\gamma$ is a function interpolating the output of an FDTD solution to Maxwell's equations (see Sec.~\ref{sec:edge_diffraction}).

\subsection{Polarization}
The initial field in the greypixel map depends on the incident polarization state relative to the local edge, therefore we must rotate into the edge frame for each cell to determine the relative contribution of $s$- and $p$-polarized light. We assume the polarization of the incident light (in the lab frame) is
\begin{equation}
    \vec{U}_0 = A\hat{\xi} + B\hat{\eta} \,,
\end{equation}
where $A=ae^{i\phi_a},B=be^{i\phi_b}$ represent the complex values of two orthogonal waves whose linear combination create the incident wave (normalized to $\left|U_0\right| = 1$). For linearly polarized light, $\phi_a=\phi_b$; for circularly polarized light, $a=b=1/\sqrt{2}$ and $\phi_a = \phi_b =\pm \pi/2$; for elliptically polarized light, $a\neq b$ and $\phi_a\neq\phi_b$.

If the normal angle of the edge (relative to vertical in the lab frame) is given by $\alpha$, and we assume horizontally polarized incident light, the polarized field diffracted by the edge (in the lab frame) is given by
\begin{equation}
    \vec{U} = R^T_\alpha\left(P_s\hat{\xi} + P_p\hat{\eta}\right)R_\alpha\vec{U}_0\,,
\end{equation}
where $R_\alpha$ is the two-dimensional rotation matrix of angle $\alpha$, which rotates from the lab frame to the edge frame. Completing the matrix multiplication, the initial field components in the lab frame are
\begin{equation}
    \begin{split}
         U_\xi  &= AM_H + BM_X \,,\\
         U_\eta &= AM_X + BM_V \,,
    \end{split}
\end{equation}
where
\begin{equation}
    \begin{split}
         M_H &= P_s\cos^2{\alpha} + P_p\sin^2{\alpha} \,,\\
         M_V &= P_s\sin^2{\alpha} + P_p\cos^2{\alpha} \,,\\
         M_X &= \sin{\alpha}\cos{\alpha}\left(P_p - P_s\right) \,.
    \end{split}
    \label{eq:B_greypixel}
\end{equation}

The matrices $M_{H,V,X}$ are the initial field maps to be applied in the Fourier transform and allows any polarization state to be applied after computing four Fourier transforms (the three in Eq.~(\ref{eq:B_greypixel}) and one for the scalar Kirchhoff field). By construction, the input polarization is horizontal, so $M_H$ represents the primary contribution from $s$-polarization, $M_V$ is the primary contribution from $p$-polarization, and $M_X$ is the cross-term due to induced polarization from the difference between the responses to $s$- and $p$-polarization. The scalar Kirchhoff field is given by
\begin{equation}
    M_K = P_K\,.
    \label{eq:K_greypixel}
\end{equation}

For a calculation with only scalar diffraction, $M_H=M_V=M_X = 0$.

\subsection{Propagation of total field}
The electric field propagated to the observation point (in the observation frame), is given by
\begin{equation}
    \begin{split}
        U_X &= C\left[A\left(F_K + F_H\right) + BF_X\right] \,, \\
        U_Y &= C\left[B\left(F_K + F_V\right) + AF_X\right] \,,
    \end{split}
    \label{eq:total_field}
\end{equation}
where the function $F$ is the Fourier transform of the initial field maps $\sigma=\left\{K,H,V,X\right\}$ times the Fresnel kernel:
\begin{equation}
    F_\sigma = \mathcal{F}\left\{M_\sigma e^{\frac{ik}{2z}\left(\xi^2 + \eta^2\right)}\right\}_{\left(x,y\right)} \,,
    \label{eq:fourier_map}
\end{equation}
and the leading constant is given by
\begin{equation}
    C = \frac{e^{ikz}}{i\lambda z} e^{\frac{ik}{2z}\left(x^2+y^2\right)} \,.
\end{equation}

The Fourier transform in Eq.~(\ref{eq:fourier_map}) is calculated with the matrix Fourier transform algorithm\cite{Soummer_2007}, which provides efficient computation when the input and output planes have largely different sampling requirements. If observed without a polarizing element at the observation point, the intensity is
\begin{equation}
    I = \left|U_X\right|^2 + \left|U_Y\right|^2 \,.
    \label{eq:unpolarized_intensity}
\end{equation}

If observed with a polarized analyzer with angle $\beta$ in the lab frame, the intensity is given by
\begin{equation}
    I = \left|U_X\cos{\beta} + U_Y\sin{\beta}\right|^2 \,.
    \label{eq:polarized_intensity}
\end{equation}

\subsection{Summary of diffraction calculation}
\begin{enumerate}
    \item For each cell in the seam, calculate the edge normal angle $\alpha$ and $P_{K,s,p}$ using Eq.~(\ref{eq:greypixel_value}).
    \item Generate four initial field maps $M_{K,H,V,X}$ using Eq.~(\ref{eq:B_greypixel}) and Eq.~(\ref{eq:K_greypixel}).
    \item Propagate maps to the observation plane with Fourier transforms via Eq.~(\ref{eq:fourier_map}).
    \item Use the incident polarization state components $A,B$ to generate the total field with Eq.~(\ref{eq:total_field}).
    \item Apply polarized analyzer to calculate the intensity via Eq.~(\ref{eq:unpolarized_intensity}) or Eq.~(\ref{eq:polarized_intensity}).
\end{enumerate}

\subsection{Numerical considerations}
The Fourier optics approach using the greypixel approximation for the Kirchhoff solution has been shown to be in agreement with accurate edge diffraction algorithms\cite{Harness_2018}. For the starshades modeled in Sec.~\ref{sec:experimental}, the scalar solution agrees with the edge diffraction algorithms with a grid size of $2^{13}\times2^{13}$. The non-scalar solutions are numerically converged with a $2^{13}\times2^{13}$ grid size, $N=100$ in Eq.~(\ref{eq:greypixel_value}), and a total seam width of 10$~\upmu$m.

\section{Calculation of boundary values in Braunbek seam}
\label{sec:edge_diffraction}
The Braunbek method requires us to know the exact field in the seam near the edge of the screen ($f_\gamma$ in Eq.~(\ref{eq:greypixel_value})). For a scalar-only solution, the field is given by the Heaviside step function as in Eq.~(\ref{eq:heaviside}). For a perfectly conducting thin screen, the field is given by Sommerfeld's solution\cite{Sommerfeld, Born_Wolf}. For more complicated materials and geometries, we solve for the field past the screen using an FDTD solution to Maxwell's equations with the open-source software \emph{Meep}\cite{Meep}, and $f_\gamma$ is the interpolation of the results as a function of distance from the edge.

\begin{figure}[htb]
\centering\includegraphics[width=0.97\textwidth]{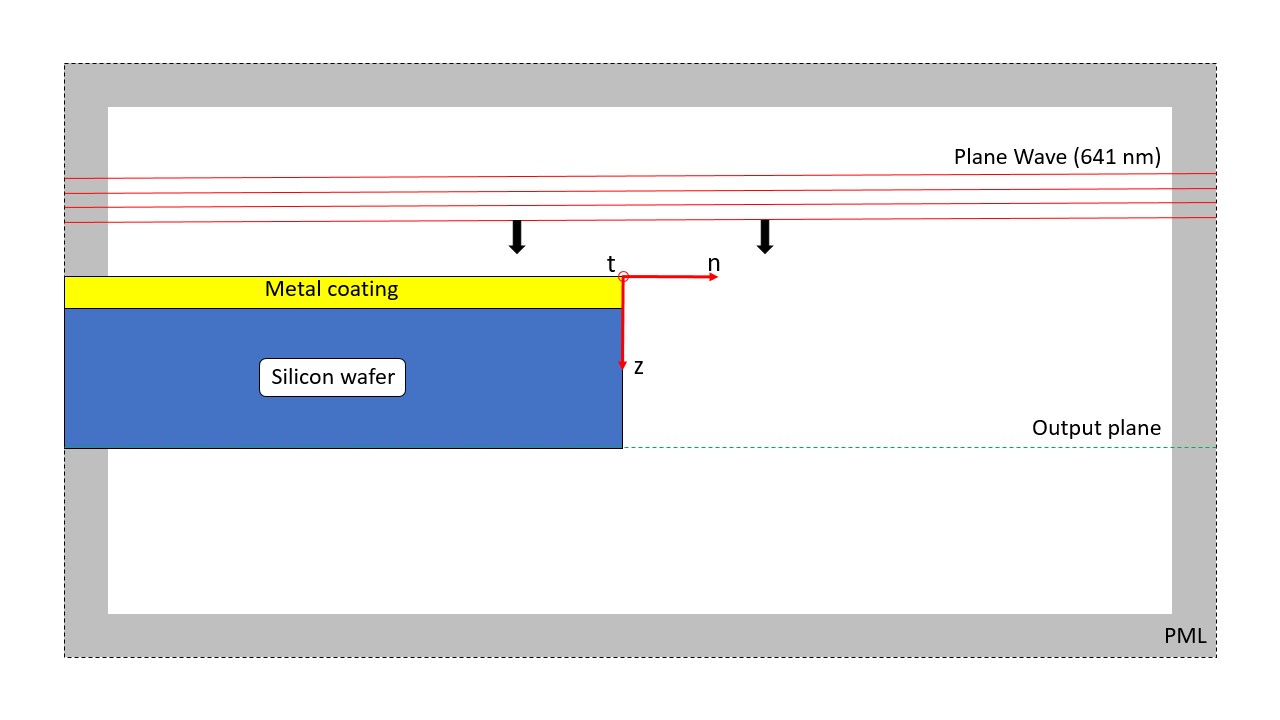}
\caption{ \label{fig:meep_cross_section}
 Cartoon diagram of the computational cell geometry for the \emph{Meep}\cite{Meep} FDTD simulation. The cell is padded by perfectly matched layer (PML) boundary conditions. The resultant field is extracted at the output plane (green dashed line) at the bottom of the wafer. The horizontal extent of the cell is truncated for clarity.
}
\end{figure}

The edges of the sub-scale starshades tested in Ref.~\citenum{Harness_2020} are comprised of a silicon wafer coated with a thin layer of metal. Figure~\ref{fig:meep_cross_section} shows a cartoon diagram of the computational cell for an FDTD simulation of light propagating past an edge. We assume the curvature of the edge in the plane of the screen is much larger than the wavelength, so that the edge can be treated as an infinite half-plane extending in the $\hat{t}$-direction, reducing the problem to two dimensions. In the simulation, we build an edge with a silicon wafer and metal coating and propagate a plane wave past the edge; perfectly matched layer (PML) boundary conditions surround the cell to eliminate numerical reflections at the boundaries. We run the simulation until the field stabilizes to a converged value and record the field at the bottom of the edge. A similar simulation is run without any edge materials to generate the incident field. Subtracting the results of the two simulations yields the additive field $(\delta)$ that is the boundary value in the Braunbek seam. These simulations are run for each polarization ($s,p$) and both the electric and magnetic fields are saved to provide the boundary values of Table~\ref{tab:delta_values}.

\begin{figure}[htb]
\centering\includegraphics[width=\textwidth]{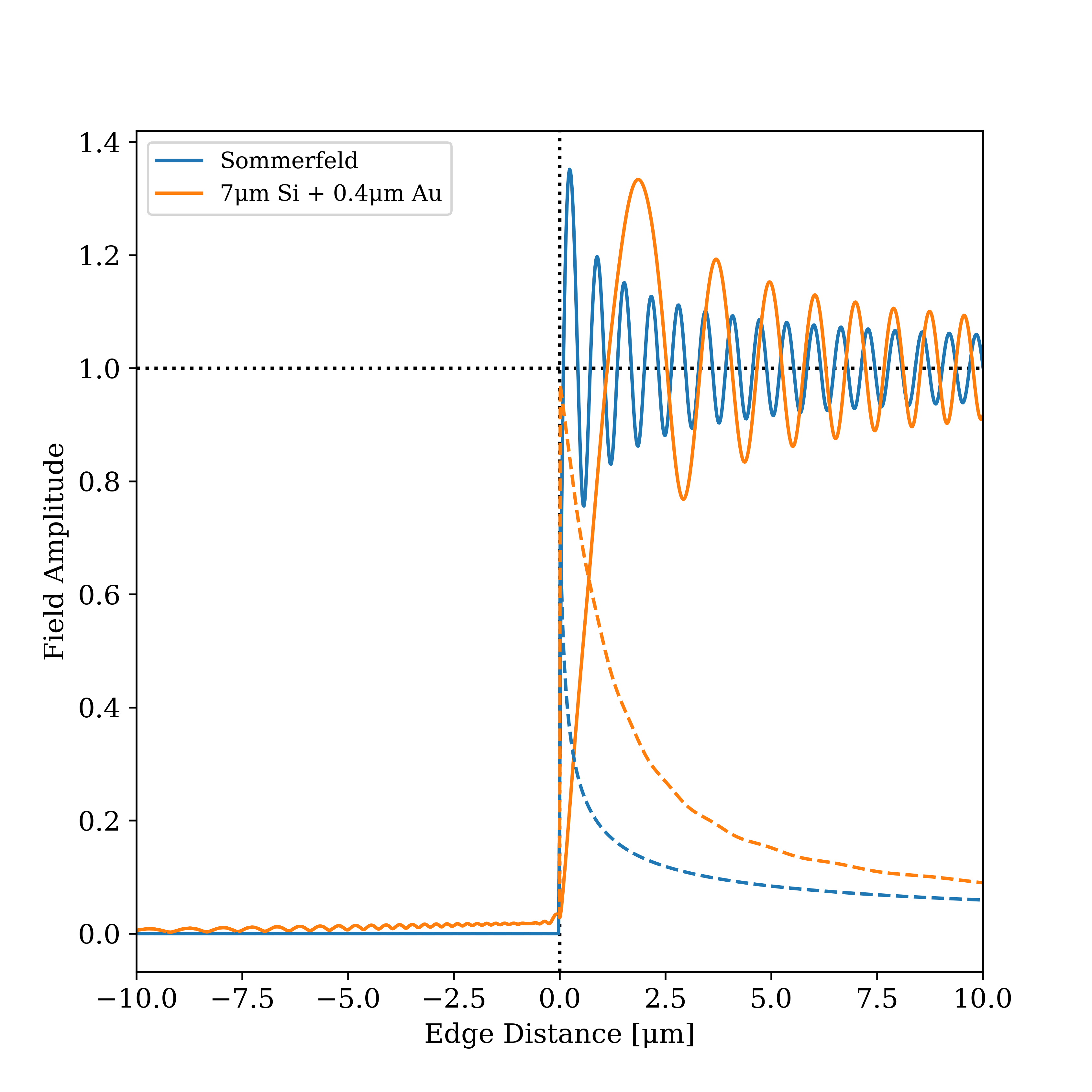}
\caption{ \label{fig:edge_fields}
 FDTD simulation results for $s$-polarized light. The solid lines are the electric fields at the bottom of the screen for simulations of a 7\micron silicon wafer with a 0.4\micron gold coating and of a thin, perfectly conducting edge (the Sommerfeld solution). The dashed lines are the additive fields $(\delta)$ calculated by subtracting the incident field. On the screen side (edge distance $<0$), the solid and dashed lines overlap as there is no incident field on the screen.
}
\end{figure}

Figure~\ref{fig:edge_fields} shows the resultant electric fields for an $s$-polarized simulation of a 7\micron silicon wafer with a 0.4\micron gold coating, and for a thin, perfectly conducting edge (the Sommerfeld solution). Also shown is the additive field $(\delta)$ calculated by subtracting the incident field. We can see that the thicker screen has a stronger impact on the electric field and that the additive field is larger in the seam of the aperture, which leads to a larger thick screen effect. In generating the greypixel map, the complex field as a distance from the edge ($f_\gamma$ in Eq.~(\ref{eq:greypixel_value})) is provided by numerically interpolating the additive field in Fig.~\ref{fig:edge_fields} as a function of distance from the edge.

The simulation geometry shown in Fig.~\ref{fig:meep_cross_section} assumes a half-plane extending infinitely behind the edge and free space in front of the edge. For the inner gaps between the petals of the starshades tested in the laboratory\cite{Harness_2020}, the gaps are only 7\micron - 16\micron across and must be simulated as a slit rather than a half-plane. For $600$ nm light, the results of a slit and half-plane converge for slit widths $>30~\upmu$m. For sharp corners in the screen, we could run an FDTD simulation of the corner and use those results for the surrounding pixels, but the total area affected by the corners is small enough to ignore this effect.

\begin{figure}[htb]
\centering\includegraphics[width=\textwidth]{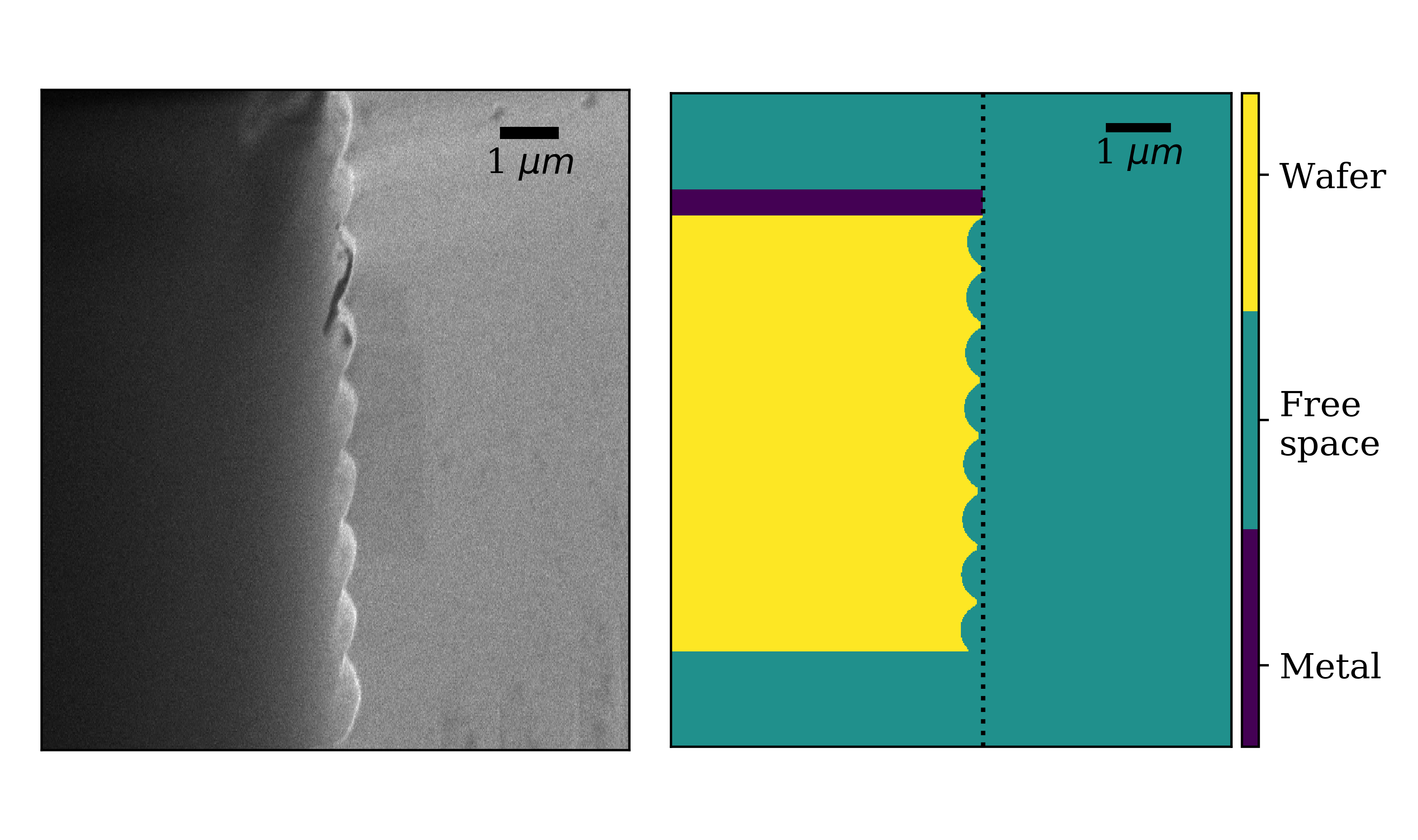}
\caption{ \label{fig:sem_image}
  Left: SEM image of the vertical profile of the wafer edge for a manufactured mask. The edge scalloping is a result of the Bosch etching process (figure provided courtesy of K. Yee, Jet Propulsion Laboratory, California Institute of Technology). Right: electric permittivity map of the \emph{Meep} FDTD simulation cell displaying the material geometry. In the simulation, we can replicate the scalloping of the edge and taper the vertical wall.
}
\end{figure}

The etching process of the starshade masks results in an optical edge that is neither flat nor vertical, both of which affect how light interacts with the edge and needs to be accounted for in the FDTD model. We are unable to generate model results that match experimental data if we assume a flat vertical edge, but a scalloped and tapered edge produces models that agree well with the data (see Sec.~\ref{sec:experimental}). In the scanning electron microscope (SEM) image of Fig.~\ref{fig:sem_image}, we can see the Bosch etching process ``scallops'' the vertical wall of the wafer and leaves divots $\sim$~0.2\micron deep and $\sim$~0.8\micron tall. The etching process can also result in a non-vertical wall, though it is more difficult to extract the wall's tapering angle from SEM images. In Sec.~\ref{sec:experimental}, we detail the analysis used to find the taper angle that best matches the experimental data. Figure~\ref{fig:sem_image} shows the simulated vertical edge profile using scallop parameters inferred from the SEM image and a 1$^\circ$ taper angle that best fits the data.

\section{Experimental validation}
\label{sec:experimental}
The sub-scale starshade experiments presented in Ref.~\citenum{Harness_2020} provide an excellent opportunity to validate our model because of the large dynamic range provided by the starshade ($10^{-10}$ in intensity) and the sensitivity of the performance to the starshade's shape. Fractional changes in the field propagating past an edge leads to order of magnitude changes in the resulting contrast. In this section, we briefly summarize results presented in Ref.~\citenum{Harness_2020} to provide experimental validation of our approach.

\begin{figure}[htb]
\centering\includegraphics[width=\textwidth]{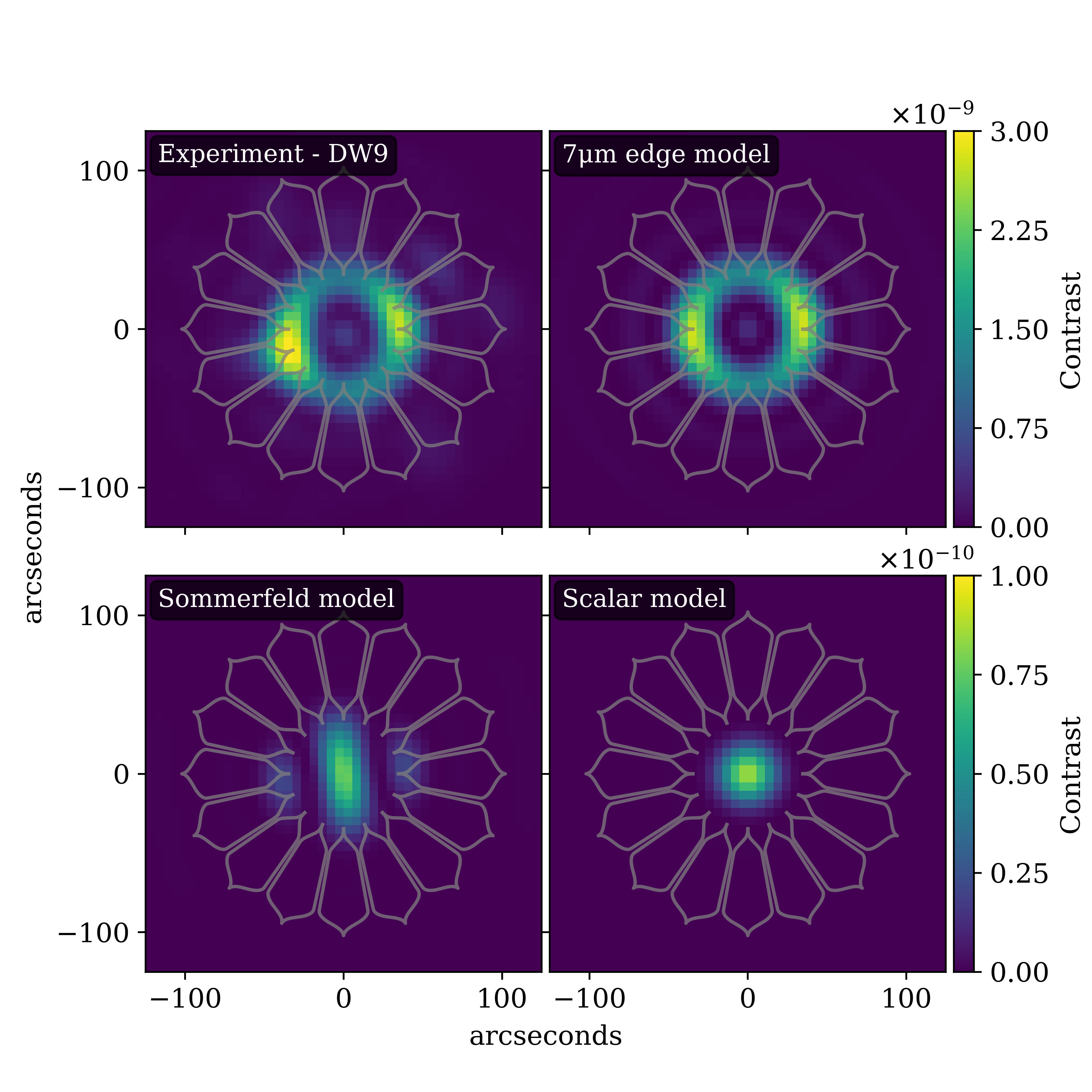}
\caption{ \label{fig:lab_v_models}
Experimental (top left) and model images of mask DW9 with starshade pattern overlaid. Top right model uses the results from an FDTD simulation of a 7\micron manufactured edge. Bottom left model uses Sommerfeld's solution for the edge. Bottom right model is scalar diffraction only. In each model the analyzer is rotated 15$^\circ$ (from horizontal) to be consistent with the experiment. The input polarization is horizontal in the image. {\bf Note} the order of magnitude difference in colorbar range between the top and bottom rows.
}
\end{figure}

The experiment design is simple: a 641 nm laser source with a diverging beam, a 25~mm diameter starshade, and a lens imaging onto a detector sit aligned in an 80~m long enclosed testbed\cite{Harness_2020}. From within the deep shadow produced by the starshade, we measure the efficiency with which the starshade suppresses the on-axis light. The laser light is linearly polarized and an analyzing polarizer on a rotation stage in front of the detector allows us to measure the polarized response of the starshade. We present results from experiments testing two different starshade masks, both etched out of a silicon wafer coated with 0.4\micron of gold. Mask DW9 has a 7\micron thick silicon edge with 7.5\micron wide gaps between petals. Mask DW21 has a 3\micron thick silicon edge with 16.2\micron wide gaps between petals. The apodization profile for the two designs are different. Note that the starshade design (seen in Fig.~\ref{fig:lab_v_models}) consists of an inner starshade that is held to the silicon wafer by radial struts, and that the outer supporting wafer is also apodized. This results in 16 individual apertures formed by the inner starshade petals, radial struts, and outer apodization. More details can be found in Ref.~\citenum{Harness_2020}.

\begin{figure}[htb]
\centering\includegraphics[width=\textwidth]{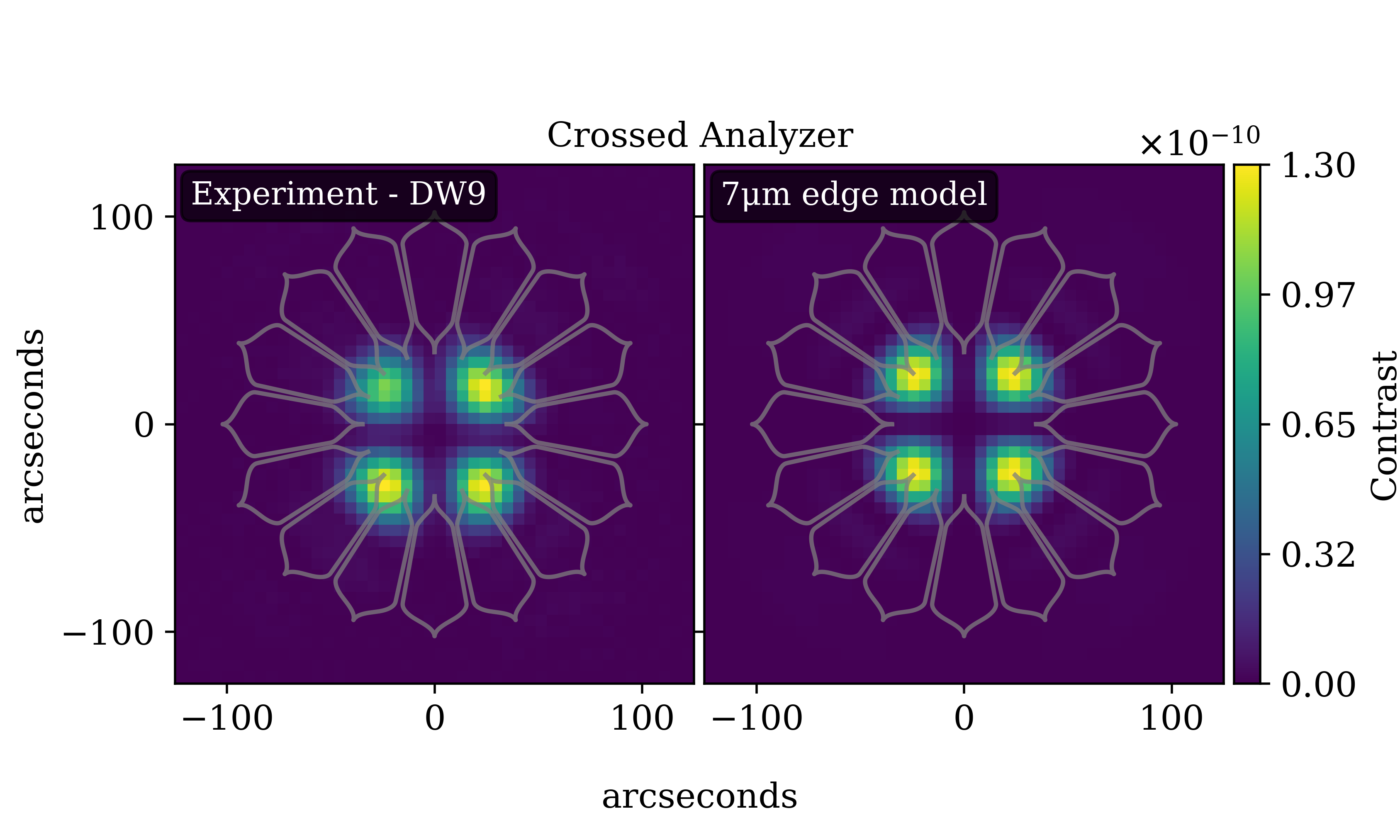}
\caption{ \label{fig:crossed_analyzer}
Experimental (left) and model (right) images of mask DW9. The camera analyzer is crossed (aligned vertical in the image) with the input polarizer (horizontal in the image); light of this polarization is solely due to the thick screen effect.
}
\end{figure}

Figure~\ref{fig:lab_v_models} shows an experimental image testing DW9 at $\lambda$ = 641 nm with the analyzer nearly aligned with the horizontal input polarization vector (there is a slight 15$^\circ$ misalignment of the analyzer). The bright lobes along the polarization axis stand out at $3\times10^{-9}$ contrast. Also shown in Fig.~\ref{fig:lab_v_models} are three models that use different field functions ($f_\gamma$ in Eq.~(\ref{eq:greypixel_value})) to generate the greypixel map. In these models, we assume the seam around the edge is 10\micron wide. The top right model uses the field function from an FDTD simulation of a 7\micron thick silicon wafer coated with a 0.4\micron gold layer, and with an edge taper angle and scalloping profile as described in Sec.~\ref{sec:edge_diffraction}. The bottom left model uses the Sommerfeld solution and the bottom right model uses the scalar only Kirchhoff solution. There is an order of magnitude difference between the experimental data and the Sommerfeld and Kirchhoff solutions, while the 7\micron edge model matches well. Figure~\ref{fig:crossed_analyzer} shows an image of DW9 with the camera analyzer orthogonal to the input polarizer, along with the 7\micron edge model. Light in this polarization is due solely to polarization induced by the thick screen effect. We see that our model matches the experimental data in two orthogonal polarization angles. Even with the complex geometries of the manufactured edge, the agreement is better than 10$^{-10}$ contrast, which we believe highlights the power of our approach.

\begin{figure}[htb]
\centering\includegraphics[width=\textwidth]{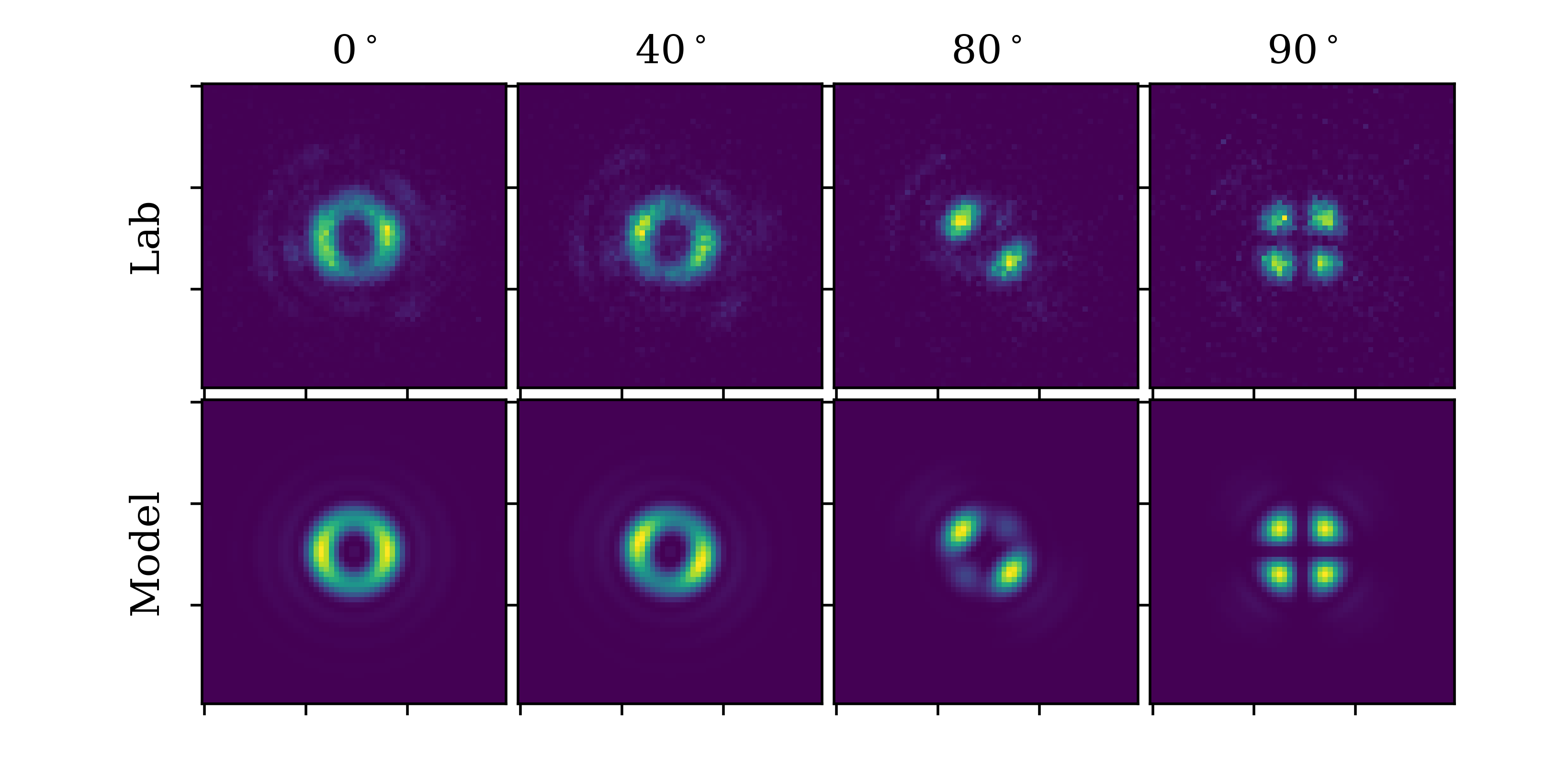}
\caption{ \label{fig:polarization_rot}
Experimental (top) and model (bottom) images of mask DW9 imaged with the camera analyzer rotating relative to the input polarizer direction (horizontal in the image). The angle given is that between the analyzer and polarizer, where 0$^\circ$ means they are aligned, 90$^\circ$ means they are crossed. Exposure times are increased for larger angles.
}
\end{figure}

The observed morphology of the polarization lobes agree with our understanding of their source. The boundary layer caused by the interaction with the edge occupies a significant fraction of the aperture only in the inner gaps between petals and thus the degradation of contrast is dominant there. The strength of the edge interaction is dependent on the polarization relative to the sidewall of the edge; $s$-polarization, where the electric field is parallel to the wall, has a greater effect as the electric field goes to zero on a perfect conductor. The inner gaps can be thought of as parallel plates aligned with the petal. If we assume horizontally polarized light, petals at the 3:00 and 9:00 positions have all $s$-polarization and petals at the 6:00 and 12:00 positions have all $p$-polarization. Since $s$-polarization has a stronger interaction, the bright lobes are aligned with the petals that have their walls parallel to the input polarization. For petals in between those positions, with walls at an angle to the polarization vector, they see both $s$- and $p$-polarization and if there is an imbalance in the interaction strength of the two, they induce a component orthogonal to the input polarization. This is seen in the four lobes of Fig.~\ref{fig:crossed_analyzer}, where the analyzer (vertical in the image) is orthogonal to the polarization vector (horizontal in the image). The morphology of the lobes at various angles can be seen in Fig.~\ref{fig:polarization_rot}, where we show images of DW9 with the analyzer rotated to different angles relative to the input polarization direction. As the contrast difference between the aligned analyzer and crossed analyzer is only a factor of 30, there are not tight constraints on the extinction of the analyzer nor on the linearity of the incident polarization; the $>>$100:1 extinction of our polarizing element is sufficient.

\begin{figure}[htb]
\centering\includegraphics[width=\textwidth]{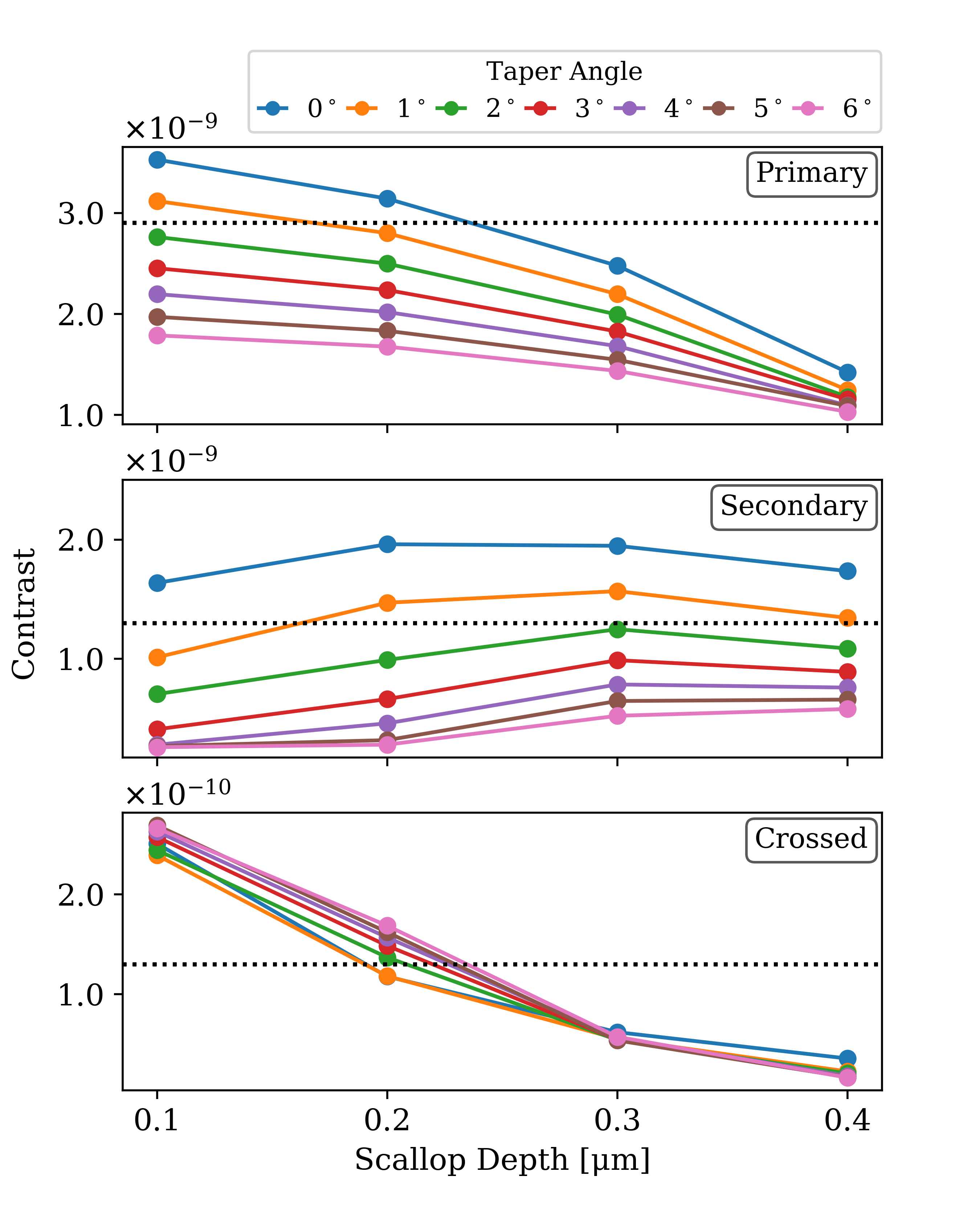}
\caption{ \label{fig:taper_trends}
Simulated contrast as a function of scallop depth and taper angle. The dotted horizontal lines are the experimental values for mask DW9, where the scallop depth and taper angle are unknown. The top plot corresponds to the primary polarization lobes, the middle plot the secondary polarization lobes, and the bottom plot the lobes in the crossed analyzer data.
}
\end{figure}

In Sec.~\ref{sec:edge_diffraction}, we mentioned that the model results do not match the experimental data unless we simulate an edge with a vertical profile that matches that of the manufactured mask. The size of the edge scallops can be inferred from an SEM image of the edge (see Fig.~\ref{fig:sem_image}), but the taper angle cannot. To match the taper angle, we ran a suite of simulations varying the angle to find which produced the best fit. The data we use to match the model with experiment are the contrasts of the primary polarization lobe (at 3:00 and 9:00 in Fig.~\ref{fig:lab_v_models}), the secondary polarization lobe (at 6:00 and 12:00 in Fig.~\ref{fig:lab_v_models}), and the lobes in the crossed analyzer data (four equal brightness lobes in Fig.~\ref{fig:crossed_analyzer}). Figure~\ref{fig:taper_trends} shows the contrast trends as a function of scallop depth and taper angle. The three experimental measurements of mask DW9 (contrast at the primary, secondary, and crossed analyzer lobes) are shown as dotted horizontal lines. The simulated data points closest to the horizontal lines yield the best estimations of the scallop depth and taper angle. The crossed analyzer contrast is a steep function of scallop depth, but does not depend on taper angle. The primary lobe contrast is moderately dependent on both, and the secondary lobe contrast is weakly coupled to scallop depth, but is dependent on taper angle. The crossed analyzer data is inconsistent with any scallop depth other than 0.2~$\upmu$m, which is consistent the value estimated from the SEM image. Given that, the best fit taper angle is 1$^\circ$, which matches well with all three data points. We sample the scallop depth and taper angle coarsely as a demonstration of the concept; more exact values could be extracted, but is beyond the scope presented here.

\begin{figure}[htb]
\centering\includegraphics[width=\textwidth]{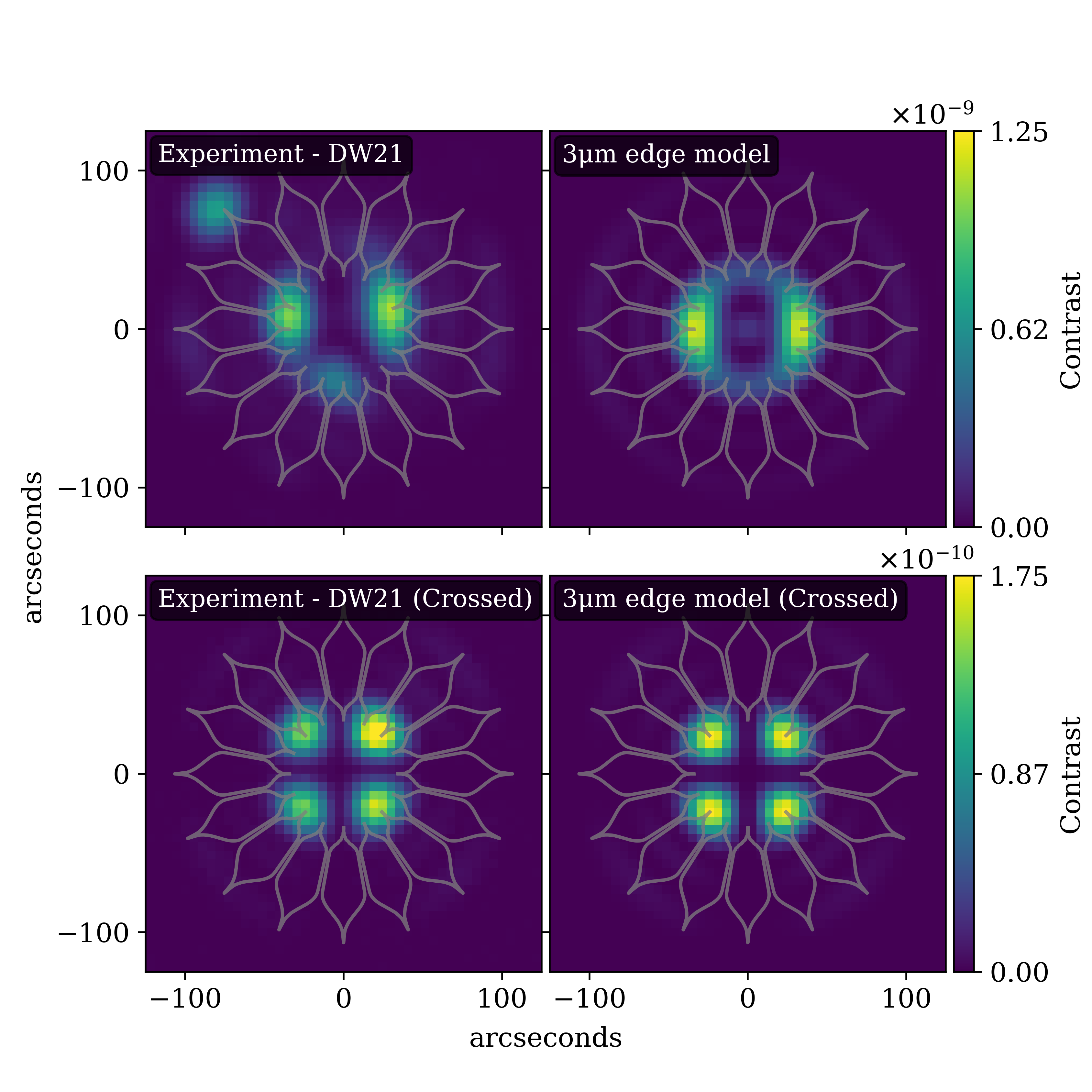}
\caption{ \label{fig:dw21_compare}
Experimental (left column) images of mask DW21 and model (right column) images of a 3\micron thick edge. In the top row, the analyzer and input polarizer are aligned, in the bottom row they are crossed. The bright spot at the 10:00 position is a manufacturing defect on the mask.
}
\end{figure}

For additional confirmation of the technique, in Fig.~\ref{fig:dw21_compare} we show the comparison between experimental and model images for the 3\micron thick mask DW21. The 3\micron thick edge model again matches the experimental data well. The best fit model has a scallop depth of 0.25\micron and taper angle of 5$^\circ$. As the gaps of this mask are wider and the edge thicker, it is reasonable to expect a different edge profile as a result of the manufacturing process. The peak contrast in the primary lobes are $\sim$ 2.3$\times$ fainter than those of DW9, which is consistent with the thick screen effect scaling linearly with edge thickness, as should be expected. We believe the lack of secondary lobe at the 12:00 position is due to interference from the manufacturing defect seen on the outer edge at the 10:00 position; the thinner mask results in a smaller degradation in contrast, which is more susceptible to interference from manufacturing defects. The dominant parameter driving the thick screen effect is the edge thickness and the resulting change in the field. Since the apodization profile is different for masks DW9 and DW21, it is not practical to compare the effect of the different gap widths between masks, however, modeling results show the change in gap widths have a small effect on the degradation in contrast for masks of the same size.

\section{Conclusions}
\label{sec:conclusions}
The method outlined in this paper provides a means to implementing non-scalar diffraction while maintaining the efficiency of Fourier optics. We derived the equations for the model that takes the complex field downstream of the edge of any diffracting element and calculates the far field diffraction for any input polarization state. We also outlined how to computationally solve for the diffraction using a greypixel map to represent the geometry of the diffracting element and its resulting edge effect. Finally, we validated the model using experimental results of sub-scale starshades and demonstrated the model replicates the observed diffraction pattern at the 10$^{-10}$ contrast level and captures the sensitivity of the thick screen effect to the edge thickness.

By restricting the area in which we deviate from scalar diffraction to a narrow seam around the edge, we only need a full electromagnetic solution over a few microns, which is readily calculated via the FDTD method. This makes the problem tractable for large optical systems that are many times the size of the wavelength of light, such as the sub-scale starshades that have features as small as 7\micron and a total extent of 50 mm. We believe this method can be used to simulate the edge effect for full-scale starshades that are 10's of meters across and to show the non-scalar effect is negligible for the large starshades. This method could also be used to include the effects of a thick mask and polarization in shaped pupil and Lyot coronagraphs\cite{Ceperley_2004}, which are also represented as binary masks and propagated via Fourier optics.

\section*{Funding}
This work was performed under subcontract with the Jet Propulsion Laboratory, California Institute of Technology under a contract with the National Aeronautics and Space Administration.

\section*{Acknowledgments}
The author would like to thank the starshade testbed team for contributions to the experiment and modeling: Stuart Shaklan, Jeremy Kasdin, Phil Willems, K. Balasubramanian, Philip Dumont, Victor White, Karl Yee, Michael Galvin, Rich Muller, Simon Vuong, and Dylan McKeithen. The author would also like to thank the anonymous referees for helpful suggestions. This project made use of the resources from the Princeton Institute for Computational Science and Engineering (PICSciE) and the Office of Information Technology's High Performance Computing Center and Visualization Laboratory at Princeton University.

\section*{Disclosures}
Experimental results are also presented in Ref.~\citenum{Harness_2020}.

\bibliography{references}

\end{document}